\documentclass[12pt]{article}
\usepackage{aaspp4}

\begin{document}
\newcommand\approxgt{\mbox{$^{>}\hspace{-0.24cm}_{\sim}$}}
\newcommand\approxlt{\mbox{$^{<}\hspace{-0.24cm}_{\sim}$}}

\def\bi#1{\hbox{\boldmath{$#1$}}}

\newcommand{\beq}{\begin{equation}}
\newcommand{\eeq}{\end{equation}}
\newcommand{\beqa}{\begin{eqnarray}}
\newcommand{\eeqa}{\end{eqnarray}}

\newcommand{\lexp}{\mathop{\langle}}
\newcommand{\rexp}{\mathop{\rangle}}
\newcommand{\rexpc}{\mathop{\rangle_c}}

\def\bi#1{\hbox{\boldmath{$#1$}}}

\title{Correlations in the Lyman $\alpha$ forest: testing the 
gravitational instability paradigm}

\author{Matias Zaldarriaga\footnote{matiasz@ias.edu}} 
\affil{Institute for Advanced Study, School of Natural Sciences,
Olden Lane, Princeton, NJ 08540}

\vskip 1pc

\author{Uros Seljak\footnote{uros@feynmann.princeton.edu}}
\affil{
Department of Physics, Jadwin Hall, Princeton University, Princeton, NJ 08544
}

\author{Lam Hui\footnote{lhui@ias.edu}}
\affil{Institute for Advanced Study, School of Natural Sciences,
Olden Lane, Princeton, NJ 08540}

\vskip 1pc


\begin{abstract}
We investigate correlations between the long wavelength fluctuations and the
small scale power in the Lyman alpha forest.
We show that such correlations can be used to discriminate between fluctuations
induced by large scale structure and those produced by
non-gravitational processes such as fluctuations in the continuum
of the quasar. The
correlations observed in Q1422+231 are in agreement with the  
predictions of numerical simulations, indicating 
that non-gravitational fluctuations 
on large scales have to be small compared those induced by the large
scale structure of the universe, contributing
less than 10 \% (95 \% confidence) of the observed power. 
We also show the sensitivity of such statistics to the equation of 
state of the gas and its mean temperature.
\end{abstract}


\clearpage

\section{Introduction}

The study of gravitational clustering in cold dark matter (CDM) models
is a well developed subject. Starting with Gaussian initial
conditions, gravitational instability can produce the large scale 
structure we observe in the universe today. Gravity, an attractive
force, tends to concentrate matter into clumps. 
This process creates non-Gaussian signatures and develops higher order
correlations. 
Through numerical N-body simulation and perturbation theory it is
possible to calculate the rate at which the density field develops
higher order moments as gravitational evolution proceeds 
(\cite{Fry94a,GGRW86,FMS93,SCFFHM98}). 
The non-Gaussianities induced by gravity satisfy
a very definite scaling with the power spectrum (or the variance
for the one point moments). When compared to the data, for example to the
distribution of galaxies in the local universe, 
this known scaling between the variance
and the higher order moments can be used to test the hypothesis that
the structure we observe has been produced by gravitational
instability from Gaussian initial conditions and that galaxies are
unbiased tracers of the mass (\cite{JBC93,Bernardeau94a,Bernardeau94b,SCO00}).
This program has lead to interesting
constraints on models for the production of initial perturbations,
ruling out some models of inflation and topological defects 
(eg. \cite{Gaztanaga94,FrGa99}). 

In the past decade a lot of progress has occurred in our understanding
of the Lyman $\alpha$ forest. By careful comparison of the data and the
numerical simulations a clear picture sometimes called fluctuating
Gunn Peterson effect, has emerged 
(\cite{chen94,her95,zhang95,m96,mucket96,wb96,th98}). 
In this picture 
most of the absorption is produced by low density
unshocked gas in the voids or mildly overdense regions in the universe. 
This gas is in ionization equilibrium and traces broadly the
distribution of the dark matter, but is also sensitive to its 
equation of state. Simple
semi-analytic models based on these ideas have been developed and shown 
to be successful in explaining the main features of numerical simulations 
(\cite{Bi92,rm95,BD97,GH96,croft97,HuiGne97,Hui97b}).  
These approximations allow one to make
simulated spectra with the correct statistical properties out of dark
matter only numerical simulations, something we will exploit in this 
paper as well. 

The realization that the main ingredient that determines the absorption
in the forest is the distribution of the dark matter has lead to the
conclusion that the forest can be a very powerful probe of
cosmology. Perhaps the most important application
is to measure the power spectrum of the dark matter at
redshifts around $z\sim 3$, which places strong constraints on the
cosmology and the nature of the dark matter
(\cite{croft98,croft99,wc2000,naray00}).  
The probability distribution of the flux and its moments has also been
computed and successfully compared with to data (\cite{McD99}). 
It has also been shown
that the moments of this distribution can be predicted analytically
using the known scalings for the matter and simple ideas of local
biasing (\cite{gaztacroft}). 

In this paper we study a particular higher order
statistic of the Ly-$\alpha$ absorption spectrum. 
Gravitational instability  predicts a
correlation between the large scale density fluctuations and the
amplitude of the density perturbations 
on small scales. Fluctuations grow faster in regions of
higher than average density, 
regardless of their wavelength (as long as it is  small 
compared to the region itself). 
This correlation between the large scale modes and the small scale
power is thus predicted to be insensitive to the amplitude 
and shape of the power spectrum.
If the primary ingredient that determines the absorption in the forest
is gravity, then this cross correlation should also be present in the
Ly-$\alpha$ flux. Thus this statistic can be used to test the general
gravitational instability scenario of the Ly-$\alpha$ forest and
constrain other processes that could influence it. As a example we
consider contamination by fluctuations in the continuum of the quasar
uncorrelated with the density fluctuations. In this paper we present
the results for this new cross correlation statistic from numerical
simulations and compare them to the values measured in quasar
Q1422+231. We use the observed values to constrain the amount of
contamination in the power spectrum on large scales.

The paper is organized as follow. In \S \ref{corr} we introduce our
statistic and in \S\ref{lyalpha} we calculate it for the flux in the
Lyman $\alpha$ forest. We also explain its dependence on the
parameters of the model. In \S \ref{Q1422} we extract this statistic
from the observed spectrum of Q1422+231. Finally, in \S
\ref{continuum} we study the effect of fluctuations in the continuum
of the quasar and use the measured
values to constrain the first of this possibilities. We conclude in \S
\ref{conclusions}.

\section{Correlations between scales in the dark
matter density field}\label{corr}

Gravity induces correlations between density perturbations on
different length scales. In particular, one can define two fields, a
smoothed density field where only the long wavelength modes are kept,
and a field that describes the spatial distribution of the power in
the small scale modes (the power here means the square of the
amplitude of the modes). This two fields will be defined rigorously
later. In this paper we focus on the correlation coefficient between
this two fields induced by the gravitational coupling between the
density modes.

\begin{figure*}
\begin{center}
\leavevmode
\epsfxsize=6.0in \epsfbox{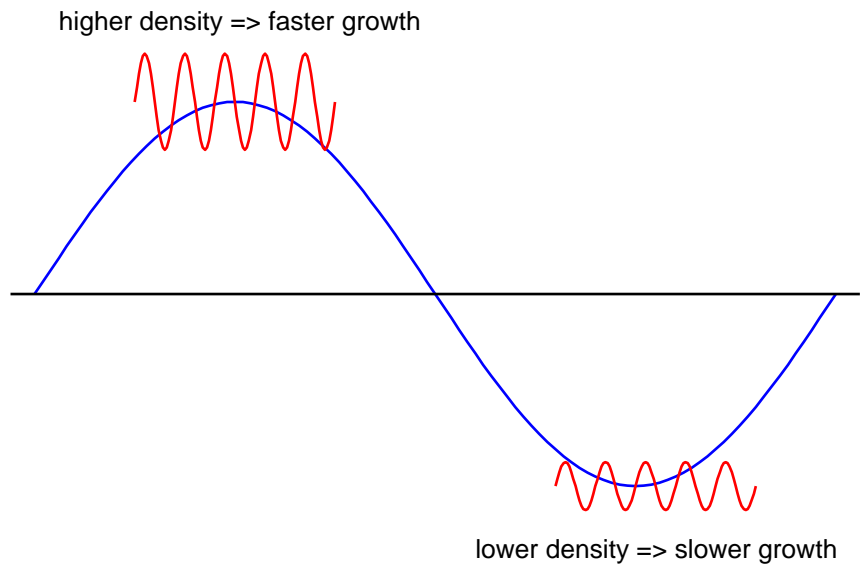}
\end{center}
\caption{The fluctuating density field has both long and short
wavelength modes. The short wavelength modes in an overdense region
caused by a long wavelength mode effectively evolve as if they are in a
universe with a higher mean density, hence they evolve faster. The oposite
is true for short wavelength modes in an underdense region. This
effect creates a correlation between the small scale power and the large scale
density fluctuations.}
\label{sketch}
\end{figure*}

Using N-body simulations it can be shown that the correlation
coefficient between the large scale density and the small scale power
is very close to one.  The reason why there is such a
large correlation can be understood with the help of the sketch in
figure \ref{sketch}.  Small scale modes in an overdense region caused
by a large scale mode evolve differently than those in an underdense
region. An overdense region has an effect equivalent to a universe
with a higher $\Omega_m$ and perturbations locally grow faster. Thus
in the overdense regions we expect there to be more small scale
power. The opposite is true for an underdense region.  The important
feature of this statistic is that it is very insensitive to the
cosmological parameters, since it only depends on the gravity being
the driving mechanism for the growth of the structures.

In this paper we study the cross correlation
statistic as applied to one dimensional data. Given
a field $f(x)$ we
compute
\beqa
F(k)&=& \int dx\ e^{ikx}\ f(x) \nonumber \\
F_{H}(k)&=& F(k)\ W_{k_1k_2}(k) 
\eeqa
where $F_{H}(k)$ is a high-pass filtered field. Fields in real
space are denoted with lower case letters, while their Fourier
transforms are written with upper case letters. 
For example, in real space the
high-passed filter field is denoted, $f_{H}(x)$. 
For simplicity we
chose a simple high-pass window,
\beq
W_{k_1k_2}(k)=\left\{ \begin{array}{ll}
			1 & \mbox{if $k_1<k<k_2$} \\
			0 & \mbox{otherwise,}
			\end{array}
		\right.
\eeq
but the results are quite insensitive to this choice as long as the
filter is non zero over a limited range of $k$ and the
range is sufficiently wide. 
We are interested in how the power on small scales (the square of the
amplitudes of the small scale modes) is distributed in space and
whether this distribution is correlated with the low $k$ modes of
$f$. For this purpose we transform the high passed filtered field
$F_{H}$ to obtain $f_{H}(x)$. We define 
\beqa f_H (x) &=& \int
{dk\over 2\pi} F_H (k) e^{-ikx} \nonumber \\ h(x)&=&f_H^2(x) \nonumber
\\ H(k)&=& \int dx\ e^{ikx}\ h(x).  
\eeqa 
The quantity $h(x)$ measures the amount of power in the wavelength
band we chose, the square in its definition is introduced so that
$h(x)$ indeed measures power and not just the amplitude of the small
scale modes. Translational invariance ensures that if
we had not introduced the square, the correlation between $h(x)$ and
the long wavelength modes of $f(x)$ is zero. 
Since $h(x)$ is a field in real space it allows us to study
how the small scale power depends on the position.

\begin{figure*}
\begin{center}
\leavevmode
\epsfxsize=6.0in \epsfbox{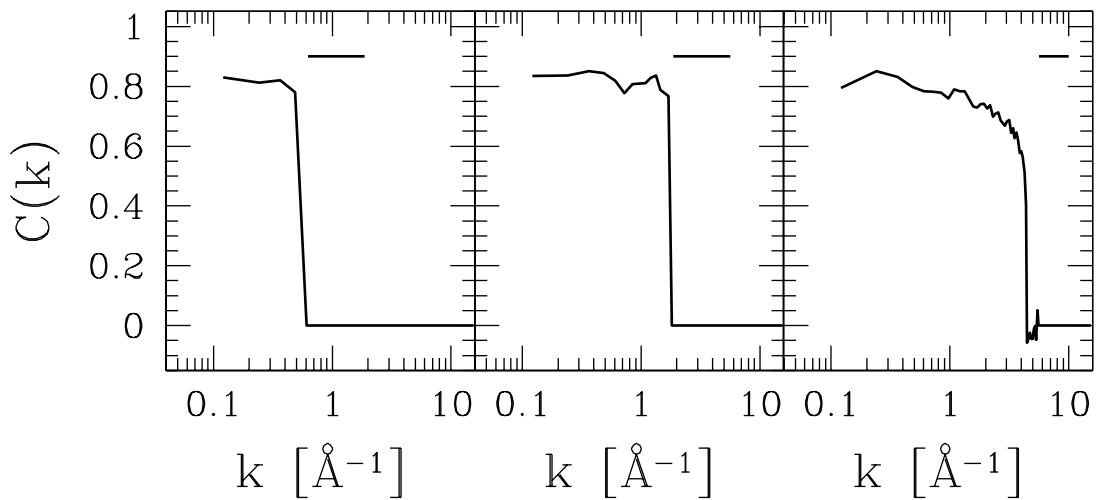}
\end{center}
\caption{The cross correlation coefficient is shown for
the density field in one dimension as defined in the text. Each panel
shows the result for a different choice of window $W_{k_1k_2}$. The
range in wavevectors used ($k_1 < k< k_2$) is shown by the horizontal
line on the top of each panel. The cross correlation has been
set to zero by hand for all wave-modes with $k > k_1$ or $k_2$, because
we are not interested in those scales (see text for details).} 
\label{ccrho}
\end{figure*}

We may further introduce the auto and cross correlation power spectra of these
quantities, as well as their cross correlation coefficient,
\beqa
\lexp F(k_1) F(k_1) \rexp &=&(2\pi)\delta^D(k_1+k_2) P^{FF}(k_1) \nonumber \\
\lexp H(k_1) H(k_1) \rexp &=&(2\pi)\delta^D(k_1+k_2) P^{HH}(k_1) \nonumber \\ 
\lexp F(k_1) H(k_1) \rexp &=&(2\pi)\delta^D(k_1+k_2) P^{FH}(k_1) \nonumber \\
C(k)&=& { P^{FH}(k) \over \sqrt{P^{FF}(k)P^{HH}(k)}},
\label{autocross}
\eeqa 
where $\lexp \rexp$ denote ensemble average, $P^{FF}$ is the
usual power spectrum of $f(x)$, $P^{HH}$ is the power spectrum of
$h(x)$ and $P^{FH}$ is the cross correlation power spectrum between
$f(x)$ and $h(x)$.  By definition the cross correlation coefficient
$C(k)$ we introduced is a number between $0$ and $1$, $C(k)=1$ means
that in both fields modes of wavelength $k$ are always identical (or
they always differ by the same multiplicative constant).  We also want
to emphasize that $h(x)$ depends quadratically on $f$ and thus
$P^{FH}$ can be expressed as a three point function and $P^{HH}$ as a
four point function of $f$.  In general these depend on the shape and
amplitude of primordial power spectrum, but for the cross-correlation
coefficient $C(k)$ this dependence is weak and $C(k)$ is always close
to unity (figure \ref{ccrho}).  In figure \ref{ccrho} we show the
cross correlation coefficient for the one dimensional density field
measured in a dark matter simulation.  We used a PM simulation
(\cite{edbert}) of the standard CDM model (SCDM) model with
$\sigma_8=0.6$. We use $16 {\rm h^{-1} Mpc}$ box with $128^3$
particles.  The three panels show the results for a different choice
of the window $W_{k_1k_2}$, the range of which is shown using a
horizontal line on the top of the figures. These ranges were chosen
for later comparison with the Lyman-$\alpha$ forest data.  We are only
interested in the cross correlation between the small and large scales
and for other wavevectors the cross correlation coefficient has been
set to zero in the plots.  We used 500 lines of sight chosen at random
from one simulation box.  The three power spectra were measured by
averaging the results from all the lines of sight and the cross
correlation coefficient was calculated by taking the ratios of the
averaged quantities.  Figure \ref{ccrho} shows that the correlation
coefficient for the 1D density is very large, approximately $C \approx
0.8-0.85$ with only a mild dependence on scale. Other cosmological
models predict very similar values for the cross-correlation
coefficient.

\section{Correlations 
between scales in the Lyman-$\alpha$ forest}\label{lyalpha}.

In the last few years a very successful model of the Lyman $\alpha$
forest has been developed.
This model, sometimes called fluctuating Gunn
Peterson effect, attributes the absorption to the smoothly distributed gas
mainly in low density regions. It has been shown that a very
simple model in which the gas is assumed to trace the dark matter
density and to be in ionization equilibrium with the
ionizing background accurately describes the relevant physics. This
model has been successfully used to compute the power spectrum of fluctuations of
the flux and the probability distribution function of the flux
(\cite{croft98,croft99}). We apply this model to PM simulations
to predict what should be expected for the cross correlation, but we
also compare it to hydrodynamical simulations (\cite{greg99}).  

In order to make predictions for this model we need to follow several
steps which have been extensively discussed in the literature
(eg. \cite{croft98}). 
The dark matter distribution is calculated with an N-body code, for which
we use a PM algorithm (\cite{edbert}). Pressure smooths the gas distribution
relative to that of the dark matter and we apply a filter
($W_F(k)=e^{-k^2/k_f^2}$) to the dark matter density to mimic this
effect. The smoothing scale $k_f$ is
related to the Jeans scale and thus to the temperature of the gas, 
but also depends on the reionization history of the gas 
(\cite{HuiGne97}). Thus
although the order of magnitude of this length scale is known, we
consider it a free parameter for the purpose of this paper
(see also the method in \cite{gh98}).
 
Once the simulation has been smoothed we
construct simulated spectra by choosing random lines of sight
in the simulation.  
We calculate the optical depth of a given fluid element as,
\beq
\tau=a_0 (\rho/\bar\rho)^\beta.
\eeq
The constant $a_0$ can be related to physical parameters (eg. \cite{croft99,Hui97b}) 
\beq
a_0=0.835 \ ({1+z \over 4})^6 \ ({\Omega_b h^2 \over 0.02})^2 \
({T_0\over 10^4\ K})^{-0.7} \ ({h\over 0.65})^{-1} \ ({H(z)/H_0\over
4.46})^{-1} \ ({\Gamma \over 10^{-12} {\rm s}^{-1}})^{-1}, 
\eeq
with $h=H_0/100 {\rm km s^{-1} Mpc^{-1}}$, $\Omega_b$ 
baryon density in units of the critical density and $\Gamma$ is the
photoionization rate. 
The position of each fluid element in velocity space is obtained by combining the
Hubble flow and the peculiar velocity of the fluid element,
$s=v_{pec}+H(z)x$, where $x$ and $s$  are the coordinates of the
element in real and redshift space, respectively, 
and $v_{pec}$ is its peculiar velocity along the line of sight.
We assume that the gas temperature is given by a simple
equation of state (eg. \cite{HuiGne97}),
\beq
T=T_0(\rho/\bar \rho)^{\alpha},
\eeq 
where $\alpha$ and $\beta$ are related by $\beta=2-0.7\ \alpha$.
Thermal broadening makes the optical depth produced by each fluid
element to be distributed in velocity space as $\exp {-(\Delta
s/b)^2}/b\sqrt{\pi}$, where $\Delta s$ is the displacement away 
from the position of the fluid element in velocity space and
$b=(2kT/m_p)^{1/2}\approx 13 \ {\rm km\ sec^{-1}} (T/10^4 {\rm K})$. 

This model of the forest is a parametrized nonlinear transformation of
the dark matter density to the flux $F=e^{-\tau}$ in velocity space
along each line of sight.  The parameters are not totally arbitrary,
since they are related to physical quantities for which approximate
values are known.  The observation of the mean transmission can be
used to fix one of the parameters, $a_0$. The mean transmission at
redshift $z\sim 3$ is observed to be $\bar F =0.684\pm0.023$
(\cite{McD99}).

In the previous section we have shown that the one dimensional density
field shows a very high correlation between the small scale power and
the large scale density modes. The model we are considering for the
observed flux in the Lyman-$\alpha$ forest is a complicated mapping
from the dark matter density field. We investigate here the expected
correlations for the observed flux. The dependence of the correlation
coefficient on the choice of the smoothing scale is very weak (as long
as the smoothing scale is taken to be independent of density). The
reason for the small dependance on $k_f$ is that if the smoothing
length is independant of density then it simply corresponds to
multiplying the Fourier components of the density field by a fixed
function of $k$. This fixed factor cancels when we take the ratio to
compute the correlation coefficient. We use $k_f=40\ {\rm h\
Mpc}^{-1}$ throughout the paper, which is close to what is expected
for reasonable reionization histories (\cite{HuiGne97}).  For each of
the values of $T_0$ and $\alpha$ we explore, we set $a_0$ so that the
generated spectra match the observed mean flux.

\begin{figure*}
\begin{center}
\leavevmode
\epsfxsize=6.0in \epsfbox{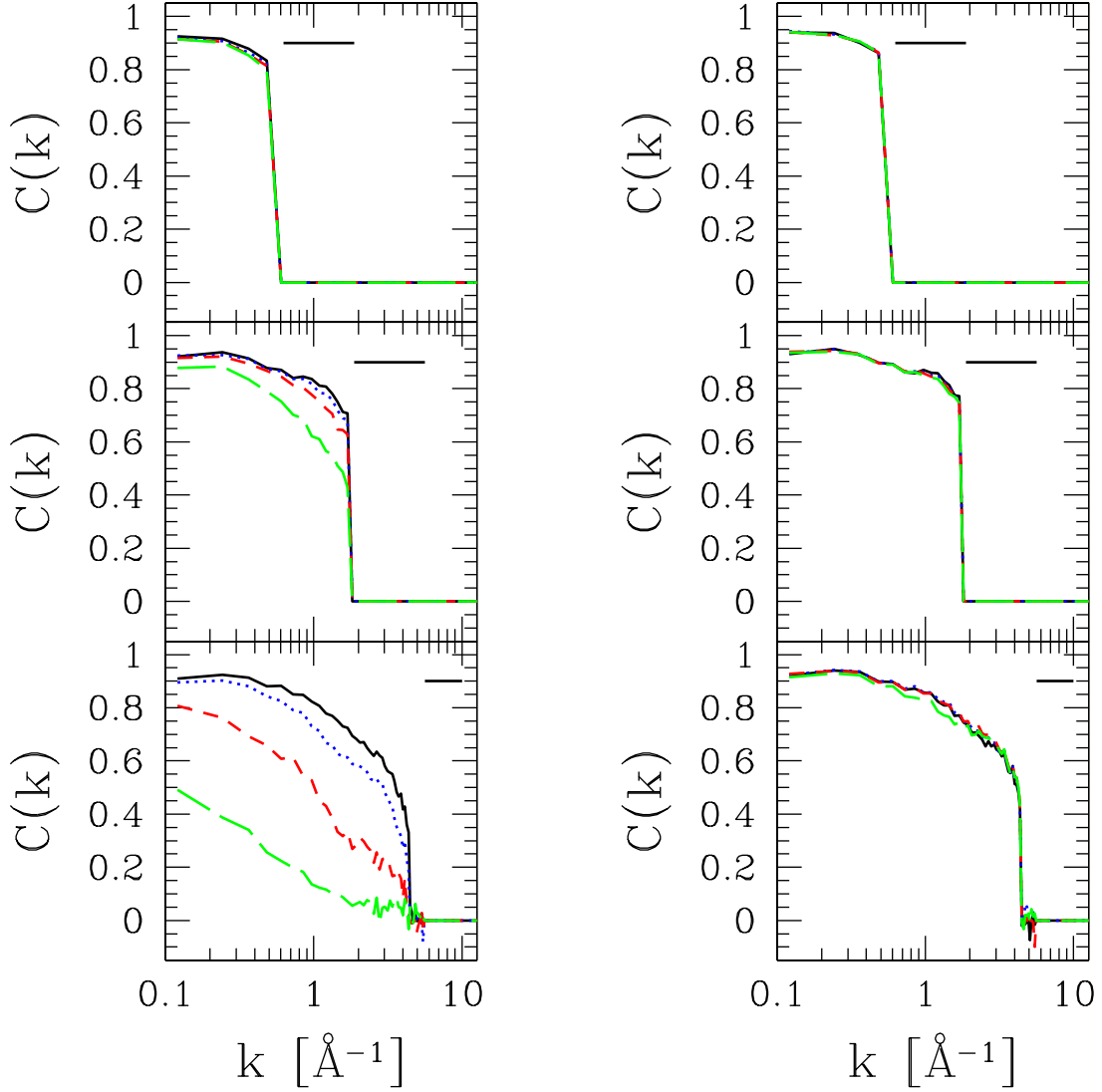}
\end{center}
\caption{Cross correlation coefficient for the optical depth $\tau$
calculated as described in the text. The left (right) columns shows the results
for $\alpha=0.6$ ($\alpha=0$). Different panels correspond to
different choices of filter $W_{k_1k_2}$. There are four lines in each
panel corresponding to different mean temperatures, $T_0=0.12,0.36,1.2,3.3 \ \times
10^{4}K$ from top to bottom in the lowest left panel.}
\label{cctau}
\end{figure*}

In figure \ref{cctau} we show the measured  
cross correlation coefficient for the optical depth $\tau$ for
different values of the temperature ($T_0=0.12,0.36,1.2,3.3 \ \times
10^{4}K$) and several choices of the high
pass filter. The left panels show the results for $\alpha=0.6$ while
the right panels show $\alpha=0$, the isothermal equation of state.
The temperatures we considered span the range allowed by other
measurements (\cite{joop99}). 
The cross correlation coefficient shown in the top panel is for
the choice of the filter 
that preserves only the largest scales (smallest $k$) and
is insensitive to the temperature regardless of $\alpha$. 
This is because these scales are larger than the
thermal broadening scale and they do not depend on the amount of
thermal smoothing. 

When we study the smaller scale filters the results
depend on $\alpha$.  For $\alpha=0$ the results are independent of
the temperature regardless of the scale. For $\alpha=0.6$ increasing 
the temperature reduces the cross correlation coefficient. 
When thermal broadening smooths the spectra with a
filter independent of density ($\alpha=0$) the
cross correlation is independent of temperature.
Thermal broadening just alters the amplitude of
the high $k$ modes, 
which cancels out when we compute the cross-correlation coefficient. Both the 
numerator and the denominator in the last line of equation
(\ref{autocross}) are changed by the same amount. 
This explains why for 
$\alpha=0$ the results show almost no change with $T_0$ 
(and also why the cross correlation is insensitive to $k_f$).  For
$\alpha>0$ the temperature depends on the density: denser
regions are hotter and thermal broadening is more important there,
suppressing the power on the smaller scales more than in colder regions. 
The higher the density,
the higher the smoothing, the lower the small scale power. This effect is the
opposite to what gravity does and thus decreases the cross
correlation coefficient. To estimate the scale at which this should 
become important we note that 
the relation between $ \AA $ 
and $\rm km /sec$ is such that
\beq
 k[ \AA ] = 61.67 ({4\over 1+z}) k[{\rm km / s}] 
\eeq
A wavenumber in the last of our bands around $k\sim 7 \mbox{\AA}^{-1}$
corresponds to $k^{-1}= 10 {\,\rm km / sec}$, which explains why temperatures
above $T=(10 {\,\rm km \ sec^{-1}})^2 \sim 
10^4 K $ significantly reduce the cross
correlation coefficient for the 
smallest scale filters when $\alpha \ne 0$. 

\begin{figure*}
\begin{center}
\leavevmode
\epsfxsize=6.0in \epsfbox{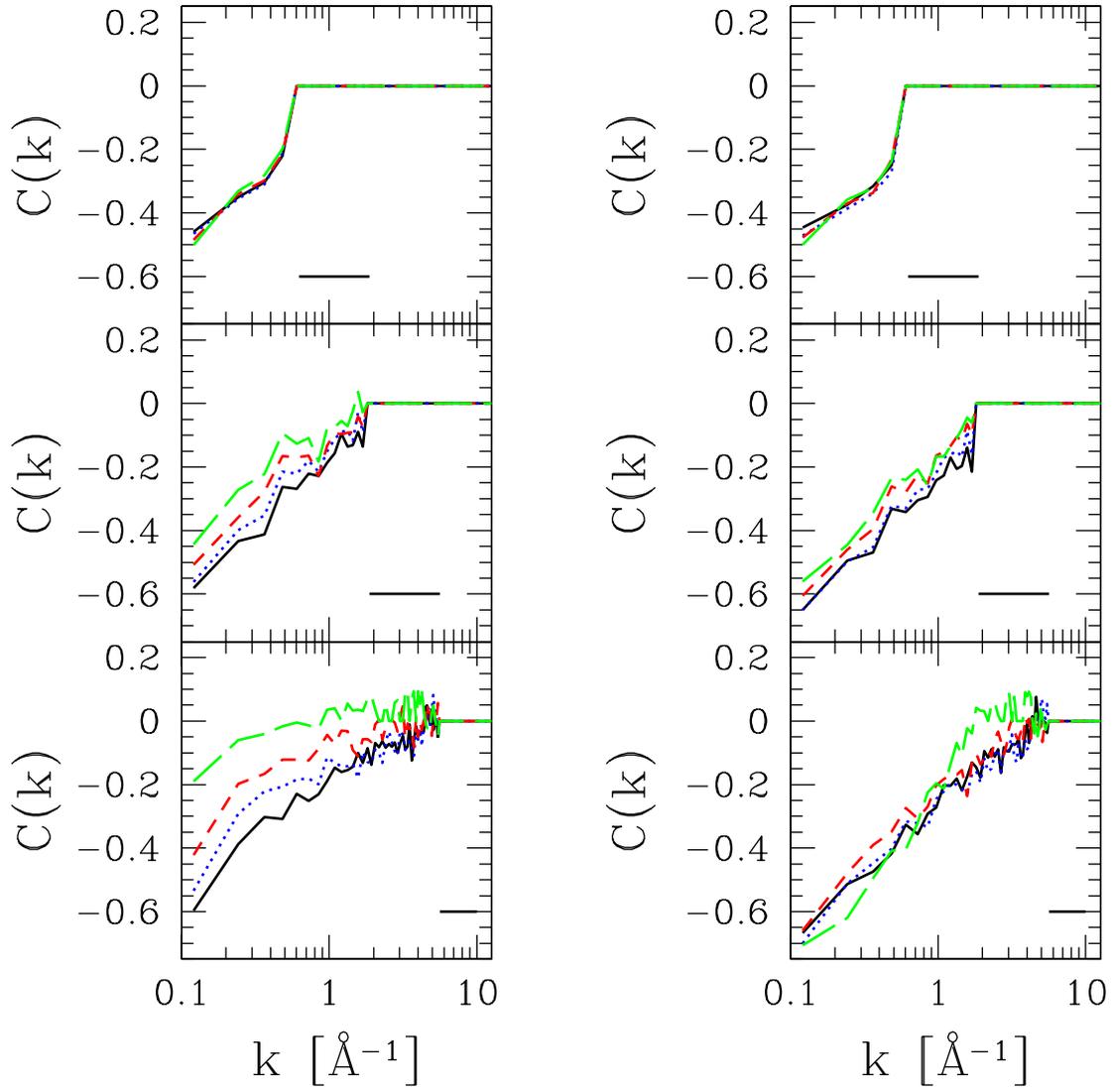}
\end{center}
\caption{Cross correlation coefficient for the flux, 
using the same parameters as in figure \ref{cctau}.}
\label{ccflux}
\end{figure*}

In figure \ref{ccflux} we show the cross correlation coefficient 
for the flux, $F=e^{-\tau}$, which can be compared directly to
the observations. 
The correlation coefficient is now negative because $e^{-\tau}$
is a decreasing function of $\tau$. It is also significantly lower,
indicating that this nonlinear transformation suppresses some of the 
correlations. It would be interesting to explore the cross-correlation
coefficient directly in the optical depth by performing the inverse 
of $F=e^{-\tau}$ on the data first, but for the purpose of this paper
we will only use the data as observed, so we do not pursue this further.
In the flux we also observe the
effect of the thermal broadening as seen in the optical depth. Again
for $\alpha=0.6$ the correlations decrease as we increase the
temperature, and the effect is more important for the higher 
bands of $k$.  

To test that our results on the cross correlation coefficient are not 
biased by using 
N-body simulations
as opposed to hydrodynamic simulations we have measured $C(k)$ in
the output of a hydrodynamic simulation for the same cosmological model 
described in Bryan et al. (1999). The results are very similar between the 
two outputs, specially on large scales ($k \,\approxlt\, 1 \AA^{-1}$).
A detailed comparison will be presented elsewhere.

\section{An application to QSO Q1422+231}\label{Q1422}


In this section we present the results of the cross-correlation 
analysis applied to the Keck HIRES observed spectrum
of Q1422+231 \cite{kim}. To measure the cross correlation 
we split the quasar spectrum between Ly$\alpha$ and
Ly$\beta$ into ten separate pieces. Each piece has a length of 3200
km/sec (16 $h^{-1}{\rm Mpc}$), the same as our simulation. 
This facilitates the
interpretation of the results and allows us to estimate the
error bars using the variance between the ten subsamples. We are
effectively considering each of the ten pieces as independent, which 
may not be a valid assumption, so we caution that the results 
presented here are preliminary. We will present a larger analysis 
elsewhere.

\begin{figure*}
\begin{center}
\leavevmode
\epsfxsize=6.0in \epsfbox{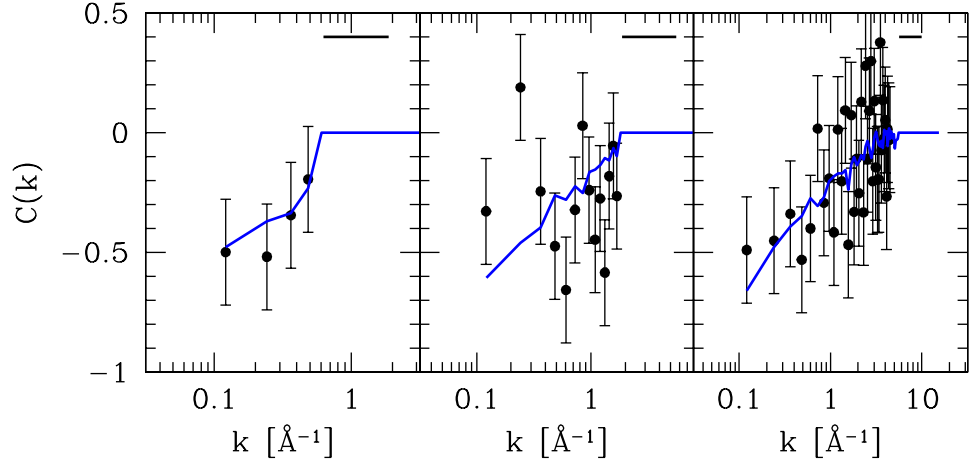}
\end{center}
\caption{Cross correlation coefficient measured in Q1422+231 together with
the prediction from simulation. The horizontal line on the top shows the range used to
construct the high-passed filter. Each point represents the mean $C(k)$
averaged over the 10 subsamples of the quasar spectra. The error bars
were computed from the simulations, but
agree with the estimate from the variance between the subsamples.}
\label{1422cc}
\end{figure*}

We measure $P^{FF}(k)$, $P^{HH}(k)$, $P^{HF}(k)$ in each of the ten
sections of the quasar spectrum. The (normalized) power spectra are measured
from the un-continuum-fitted data using the trend-removal technique
as described in Hui et al. (2000).
We choose 3 high passed
windows as used in the simulations.
We construct the cross correlation coefficient form the ratios of
these  power spectra for each of the 10 sections
independently and estimate the variance from the fluctuations between the
subsamples. Figure \ref{1422cc} shows the mean of the cross
correlation in the 10 subsamples. The errorbars computed using
the simulations agree with those inferred from the 
variance between the subsamples. It should be noted that because $C(k)$ is
a number between -1 and 1, the shape of the error distribution is not
Gaussian and the error bars in the plot should only be taken as an
indication of the variance. For comparison we also show the
values measured in the simulation for $\alpha=0$ and $T_0=1.2\ 10^4
K$. The agreement between prediction and observations is quite
striking, lending further support to the overall picture of the
forest as being generated by gravitational instability. 

\begin{figure*}
\begin{center}
\leavevmode
\epsfxsize=6.0in \epsfbox{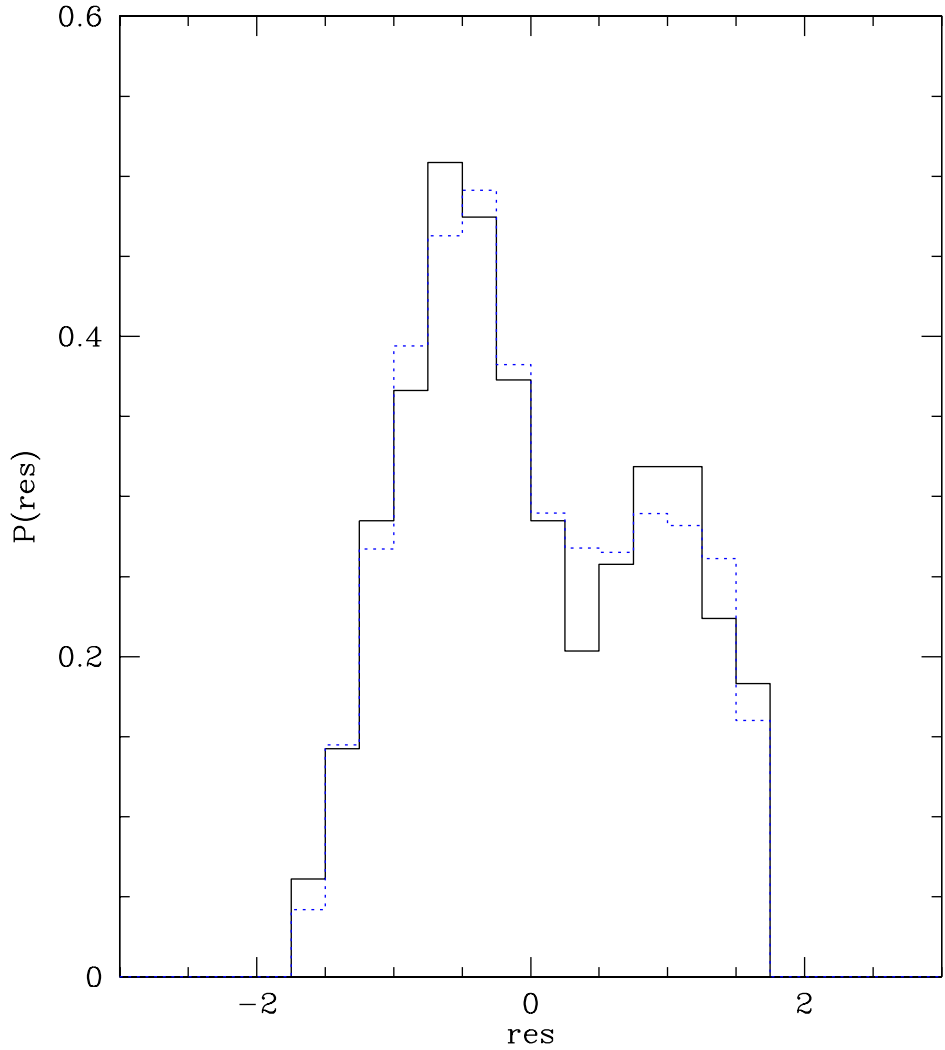}
\end{center}
\caption{The distribution function of
the residuals of the measured $C(k)$ for 
QSO1422+231 (dashed line) and simulations (solid line).
The residuals are defined as $(C-\bar C)/\sigma_C$ where $C$
is the cross correlation coefficient measured in each of the 10
subsamples  of the spectra, $\bar C$ is the average over the subsamples and
$\sigma_C^2=0.5*(\bar C^2+1)$ is a normalization factor introduced for
convenience in the plotting. 
}
\label{resc}
\end{figure*}

To further check the agreement between the observed data and the
simulations we computed the residuals from the mean of $C(k)$ for
the different subsamples of the spectra.  
Figure \ref{resc} shows the histogram of these residuals, both 
for the data and simulations. 
The distribution of residuals 
is very similar for the two samples, so 
one can use the distribution observed in
the simulations to constrain processes other than
gravitational instability that could be affecting the observed cross
correlations. 

The first important conclusion to draw from figure \ref{1422cc} is
that we observe a negative cross correlation coefficient
in the range predicted by the 
simulations. Not only are the values of the cross correlation
coefficient similar to those in our simulations but the distribution
of their residuals from the mean is also very consistent. In the next
section we use this measurement to constrain contamination from 
fluctuations in the continuum. Here we compare the smallest
scale filter (the right panel in figure \ref{1422cc}), in which we observe
significant cross correlation, with the figure \ref{ccflux}. The observed 
correlation argues against models with both high temperature $T_0$ and high
$\alpha$ (the top lines in the bottom left panel of figure
\ref{ccflux}). More accurate constraints on these parameters from such
observations will be presented elsewhere, as the details in this regime
are somewhat sensitive to other parameters like the Jeans smoothing 
scale and numerical resolution. It seems however that this can be 
a promising way to study the state of the gas at high redshifts.

\section{Testing the continuum
and the optical depth fluctuations}\label{continuum}

In previous sections we have calculated what we expect for the
cross correlation coefficient in the flux of the Lyman
$\alpha$ forest in the currently accepted model for the forest and
compared that with the measurements in Q1422+231. This
statistic can also be used as a test of the contamination from 
non-gravitationally induced fluctuations in the forest and in this
section we want to explore its discriminatory power. 

We consider the effects of the quasar continuum and optical depth 
fluctuations. We can model the observed flux as,
\beq
F(s)=[1+c(s)] e^{-\tau}
\label{fluxpcont}
\eeq 
where $c$ is a slowly varying function of position across the
spectrum. Continuum changes the overall level of the flux and so 
acts multiplicatively on the fluctuations produced by the neutral gas.
This case also applies to any process that adds to the
optical depth, such that $\tau \rightarrow \tau+\delta \tau$,
if we identify $\delta\tau=\ln(1+c)$. 

Let us consider the case where $c(s)$ has 
predominantly large scale power so that
the fluctuations in $c$ dominate over the fluctuations in $e^{-\tau}$
on large scales. If we consider a region where $c(s)>0$
the flux will
be larger than the mean flux in this region, $F(s) > \bar F$. On small
scales, the power in this region will be enhanced because the
continuum or optical depth fluctuations enter multiplicatively in 
the flux. The amplitude
of all the small scale fluctuations is enhanced by a factor $[1+c(s)]$. Thus
the continuum fluctuations induce a positive cross correlation
between the large scales and the small scale power. This is 
opposite in sign of what gravity induces and constitutes
the basis for the discriminatory power of this statistic.

\begin{figure*}
\begin{center}
\leavevmode
\epsfxsize=6.0in \epsfbox{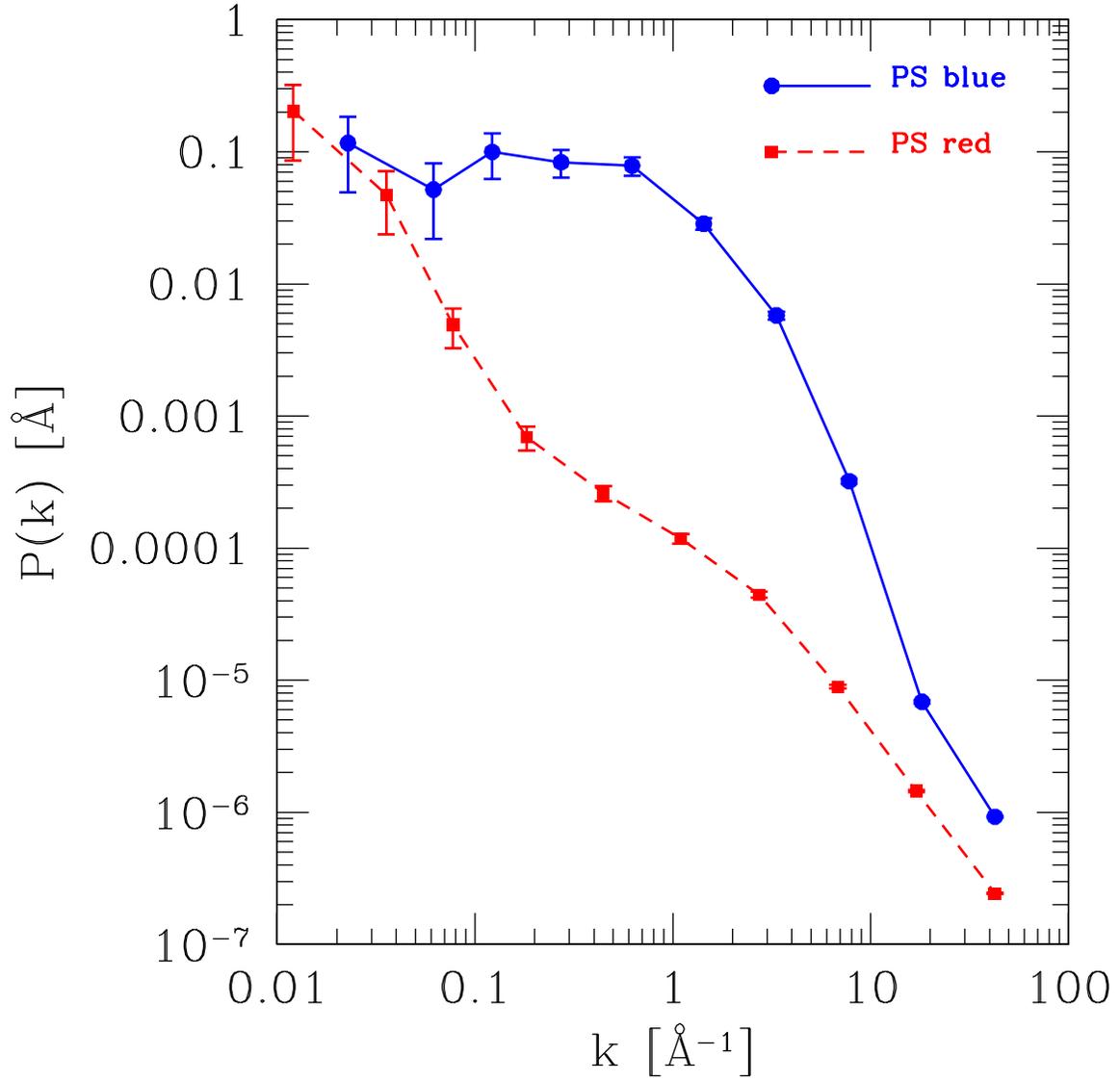}
\end{center}
\caption{Power spectrum of the fluctuations in the flux 
on the red side and blue (L$\alpha$ forest) side of the
L$\alpha$ line of of Q1422+231. }
\label{psfigure}
\end{figure*}

The detailed predictions for the effect of the continuum
will depend on its level of fluctuations compared to those induced by
the absorption as a function of scale. For
the purpose of the study here we assume that we can describe the
continuum fluctuations with a power spectrum of shape similar to that
of the Lyman $\alpha$ forest on large scales, but with an exponential
cutoff at $k \sim 1$ ~\AA\ $^{-1}$. We will let the normalization
of the additional power spectrum to be a free parameter which we try to
constrain.  
We choose this shape for the power spectrum 
of fluctuations so that it is the most degenerate
with the expected shape of the power spectrum induced by gravity, so
that the presence of continuum fluctuations cannot be revealed by a change
in the shape of the measured power spectrum. We introduced an
exponential cutoff so that 
the fluctuations are only important on large scales,
where we expect them to be the largest based on the continuum fluctuations on
the red side of Ly-$\alpha$ (figure \ref{psfigure}; see \cite{hui00}).

\begin{figure*}
\begin{center}
\leavevmode
\epsfxsize=6.0in \epsfbox{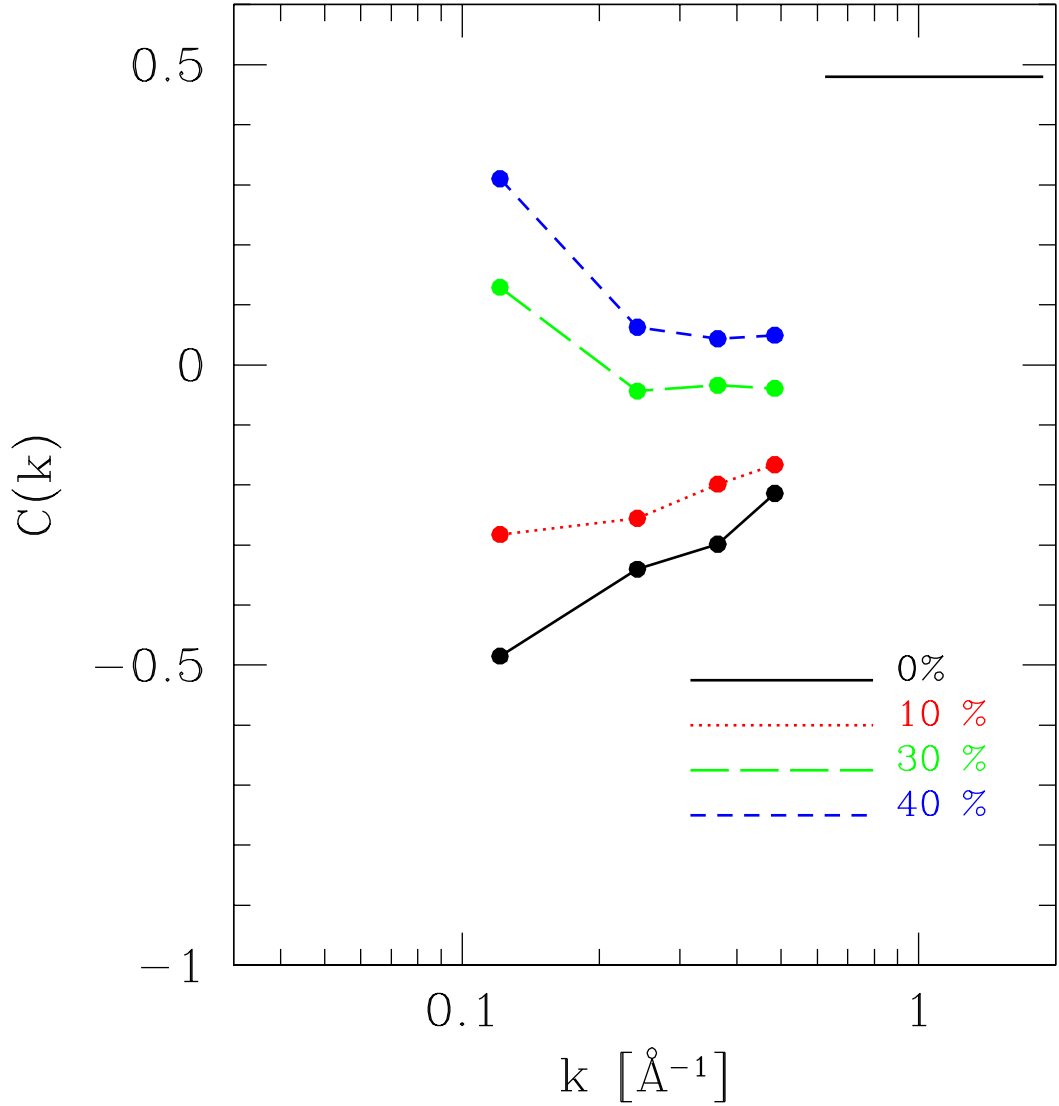}
\end{center}
\caption{Cross correlation coefficient for different choices of the
amplitude of the continuum fluctuations corresponding to 10, 30 and 40 \%
of the flux power spectrum. We also show the case of no contamination 
as a solid line.}
\label{cccont}
\end{figure*}

In figure \ref{cccont} we show the
expected cross-correlation for the lowest $k$ filter for several
choices of the amplitude of the continuum power spectrum. As expected,
increasing the amplitude drives the
correlation to positive values.  
The effect of the fluctuations of the QSO continuum 
can be used to constrain the amplitude of the fluctuations using
the measurements of Q1422+231.
To do so we focus
on the results in the lowest $k$ range. As we discussed in the previous
section these are most insensitive to other parameters in our
model and the predictions of our N-body model are in best agreement with
the results of the full hydrodynamic simulations.  

\begin{figure*}
\begin{center}
\leavevmode
\epsfxsize=6.0in \epsfbox{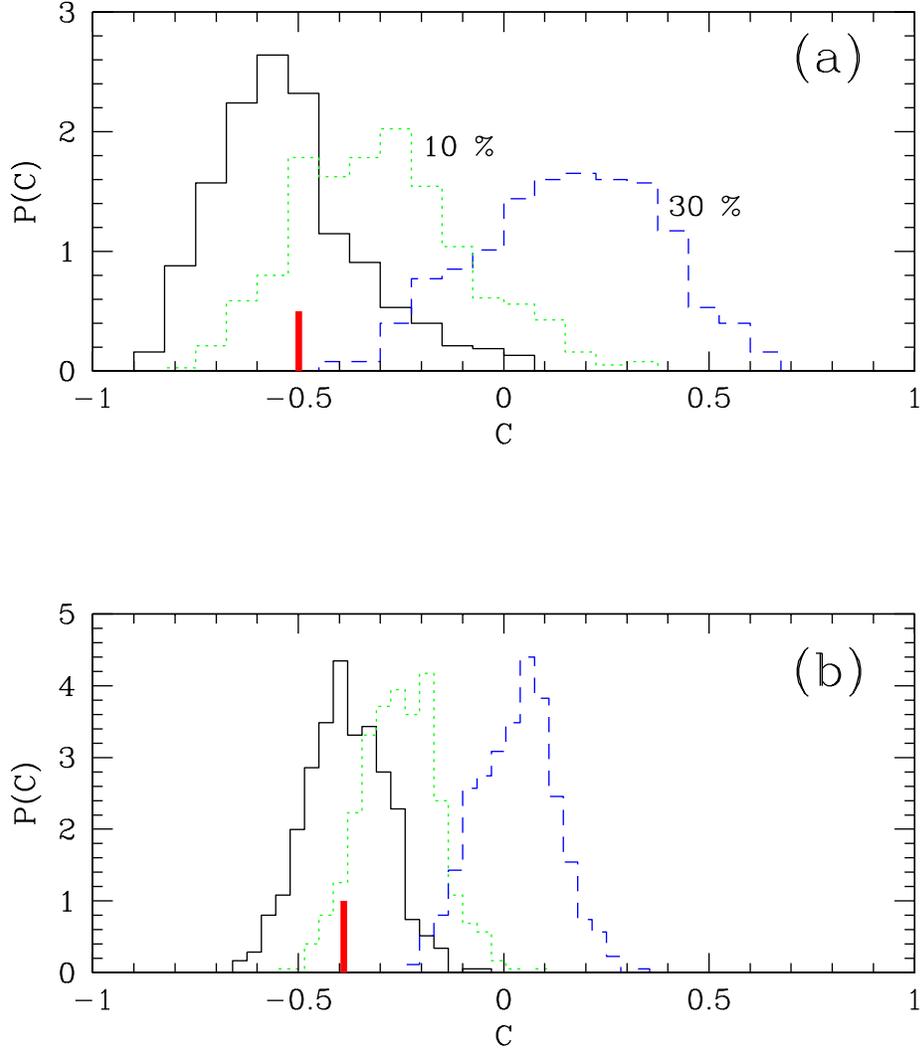}
\end{center}
\caption{Histograms of cross correlation coefficients measured in
simulations for cases when the size of continuum fluctuations is 10\% and
30 \% of the amplitude of the fluctuations in the flux. We
also show the case of no continuum fluctuations (solid line). 
Panel (a) is for the
lowest bin of our lowest high pass filter range, shown in the leftmost
panel of figure \ref{1422cc})
and panel (b) is for
the average over all 4 bins for the same filter shown
in the same panel.
To construct the histogram
we extracted 5000 lines of sight from the simulation, each line of
sight has a length of $1/10$ of the spectra of Q1422+231.
The vertical bar in each panel indicate the measurements
from Q1422+231.}
\label{histocont}
\end{figure*}

We can constrain the continuum fluctuations using a Monte Carlo
technique. We generate random spectra with different levels of continuum
contaminations to simulate the measurements of Q1422+231. 
We use realizations of Gaussian random field to simulate the continuum.
We repeat this process many times and for several levels of
contamination to measure the
distribution of values for the cross correlation coefficient that we
should observe in Q1422+231. We then 
compare these distributions with the measured values to put a
constraint on the level of contamination.  
In figure \ref{histocont} we show the distribution of measured 
cross-correlation coefficients in the simulations assuming that the
continuum fluctuations are 10 \%, 30 \% and 40 \% of the overall power of the
flux fluctuations. We also show the distribution when there is no 
contamination.
The upper panel is for $C(k=0.12)$, 
the lowest $k$ bin, and the bottom panel is for the average of the
four bins. The vertical
line on the x axis shows the measured value in Q1422+231. It is clear that
continuum fluctuations that contribute 30 \% of the measured power 
are ruled out and that even a 10 \% contamination level is unlikely. 
This is in agreement with the expected level of continuum fluctuations
based on the power spectrum of the 
red side of the Ly-$\alpha$ emission, which, except for
the largest scales (not probed here),
is significantly below the power spectrum of the blue
side dominated by the Ly-$\alpha$ absorption (figure \ref{psfigure}).
With a larger sample of QSO it will be possible to explore also 
the scale dependence of the continuum contamination. 
Cross-correlation information can also be used to subtract out the 
contamination from the power spectrum, at least if it does not 
completely dominate the fluctuations on large scales as the results 
from figure \ref{psfigure} suggest. 

We stress that as we discussed in the previous section we did not try
to correct for continuum fluctuations before measuring the Fourier
components of the flux. We directly measured the Fourier modes of the
relative flux fluctuations in each chunk of the spectrum and did the
same thing in our Montecarlo simulations. This is particularly
important for our test because one can imagine that the continuum
fitting procedure could introduce spurious correlations between scales
that under some circumnstances could mimick the effect of gravity.


\section{Conclusions}\label{conclusions}

Gravity predicts that nonlinear evolution generates higher order correlations
from an 
initially Gaussian density
field. One of the consequences
of the gravitational instability picture is that there should be a
correlation between the power on small scales and the large scale
modes of the density. This correlation is induced because in locally 
overdense regions the fluctuations grow faster, while the opposite is
true for the underdense regions. While such a cross-correlation 
only uses a subset of all the information contained in the higher 
order correlations, it has the advantage of being relatively 
insensitive to the cosmological model and the initial spectrum of 
fluctuations. This allows one to separate the effects induced by 
gravity from those induced by other effects in a model independent way.
We have presented a statistic that is sensitive
to gravity and applied it to the flux measurements in the Lyman
$\alpha$ forest.

The values of the correlation coefficient measured in Q1422+231 are in 
good agreement with what is found in the simulations. This is another
confirmation of the overall picture of the forest based on the 
gravitational instability and ionization equilibrium. The agreement
can also be used to constrain other processes that could contaminate 
large scale fluctuations in the forest and thus degrade the 
extraction of the dark matter power 
spectrum from them. If such processes unrelated to gravity 
were present they would suppress the
cross-correlations. In this paper we estimate the level of fluctuations in
the continuum of the quasar on large scales and show that even a 10\%
level of contamination on 10h$^{-1}$Mpc scales 
is unlikely, assuming the continuum has the same
power spectrum shape as the fluctuations induced by gravity.
With a larger sample of quasars we can expect to be able to determine
the level of contamination as a function of scale, which will allow one 
to extract the dark matter power spectrum to significantly larger 
scales than current estimates permit (\cite{croft99,McD99};
see also \cite{hui00}).

The cross-correlation analysis can also be used 
to set constraints
on the variation of other quantities, such as the parameters of the
equation of state of the gas. The preliminary comparison of the cross
correlation with the simulations seems to indicate that models with
both high temperature and large $\alpha$ are inconsistent with the
data. With a larger sample one should be able to determine 
with higher precision parameters 
such as the mean temperature of the gas and its 
equation of state. This will provide 
important constraints on the thermal history of the intergalactic medium at high
redshifts. 

The authors wish to thank Roman Scoccimarro and Greg Bryan for very
helpful discussions. We also thank Greg Bryan for providing us with
the results of his hydrodynamic simulations. 
We are grateful to Antoinette Songaila
and Len Cowie for kindly making available the spectrum for Q1422+231.
Support for this work
is provided by the Hubble Fellowship (M.Z.) 
HF-01116-01-98A from STScI, operated by
AURA, Inc. under NASA contract NAS5-26555, 
the NASA grant NAG5-8084 (U.S.), the NASA grant NAG5-7047,
NSF grant PHY-9513835 and the Taplin Fellowship (L.H.).

\end{document}